\documentclass[11pt, a4paper]{article}
\usepackage{jcappub}

\usepackage{booktabs}
\usepackage{multirow}
\usepackage{amssymb}
\usepackage[export]{adjustbox}
\usepackage{floatrow}
\newfloatcommand{capbtabbox}{table}[][3.5in]
\newfloatcommand{capbfigbox}{figure}[][2.3in]
\usepackage{blindtext}
\usepackage{amsmath}
\usepackage{booktabs}
\usepackage{aas_macros}

\def\lsim{\mathrel{\raise.3ex\hbox{$<$\kern-.75em\lower1ex\hbox{$\sim$}}}}
\def\gsim{\mathrel{\raise.3ex\hbox{$>$\kern-.75em\lower1ex\hbox{$\sim$}}}}

\begin{document}

\hspace*{110mm}{\large \tt FERMILAB-PUB-16-178-A}

\vskip 0.2in


\title{A Case for Radio Galaxies as the Sources of IceCube's Astrophysical Neutrino Flux}

\author{Dan Hooper}
\emailAdd{dhooper@fnal.gov}

\affiliation{Fermi National Accelerator Laboratory, Center for Particle
Astrophysics, Batavia, IL 60510}
\affiliation{University of Chicago, Department of Astronomy and Astrophysics, Chicago, IL 60637}
\affiliation{University of Chicago, Kavli Institute for Cosmological Physics, Chicago, IL 60637}

\abstract{We present an argument that radio galaxies (active galaxies with mis-aligned jets) are likely to be the primary sources of the high-energy astrophysical neutrinos observed by IceCube.  In particular, if the gamma-ray emission observed from radio galaxies is generated through the interactions of cosmic-ray protons with gas, these interactions can also produce a population of neutrinos with a flux and spectral shape similar to that measured by IceCube. We present a simple physical model in which high-energy cosmic rays are confined within the volumes of radio galaxies, where they interact with gas to generate the observed diffuse fluxes of neutrinos and gamma rays. In addition to simultaneously accounting for the observations of Fermi and IceCube, radio galaxies in this model also represent an attractive class of sources for the highest energy cosmic rays.}

\maketitle

\section{Introduction}

In 2013, the IceCube Collaboration published the first detection of high-energy astrophysical neutrinos~\cite{Aartsen:2013bka}. Subsequent analyses~\cite{Aartsen:2015knd,Aartsen:2015rwa,Aartsen:2014gkd,Aartsen:2013jdh} have found IceCube's events to follow an approximate power-law spectrum extending from tens of TeV to a few PeV, and with flavor ratios consistent with those predicted from pion decay~\cite{Aartsen:2015ivb}. Although a variety of astrophysical sources for TeV-PeV neutrinos have been proposed over the years (for reviews, see Refs.~\cite{Becker:2007sv,Halzen:2002pg}), many of these source classes now appear to be disfavored. Perhaps most notably, gamma-ray bursts had long been considered to be among the most promising sources for the ultra-high energy cosmic rays~\cite{Waxman:1995vg,Vietri:1995hs,Milgrom:1995um} and a likely source of high-energy neutrinos~\cite{Waxman:1997ti,Rachen:1998ir,Guetta:2003wi}. The lack of any detected correlations in time between IceCube's events and observed gamma-ray bursts has all but ruled out this possibility, however~\cite{Abbasi:2012zw}. Furthermore, a joint analysis of IceCube's events with data from the Fermi Gamma-Ray Space Telescope has lead to the conclusion that less than 20\% of IceCube's flux can originate from blazars~\cite{Glusenkamp:2015jca} (see also Ref.~\cite{Ahlers:2014ioa}). Similarly, a recent analysis has demonstrated that star-forming galaxies can generate no more than 28\% of IceCube's observed spectrum~\cite{Bechtol:2015uqb}. 

In light of these and other constraints, radio galaxies (active galaxies with jets that are not aligned along the line-of-sight) now appear to be, perhaps, the most promising class of sources for IceCube's observed neutrino flux~\cite{Tjus:2014dna,Giacinti:2015pya,Murase:2015ndr}. In contrast to blazars, which are generally thought to be the subset of active galaxies whose jets are directed within approximately $14^{\circ}$ of Earth~\cite{Urry:1995mg}, radio galaxies appear individually less luminous, but are much more numerous. Radio galaxies are further classified according to their morphological characteristics as either Fanaroff-Riley Type I or Type II galaxies, which are generally interpreted as the misaligned counterparts of BL Lacs and flat spectrum radio quasars, respectively.

In a recent study~\cite{Hooper:2016gjy}, it was demonstrated that the isotropic gamma-ray background (IGRB) measured by the Fermi Gamma-Ray Space Telescope~\cite{Ackermann:2014usa} is dominated by emission from unresolved radio galaxies, along with a smaller but non-negligible contribution from blazars~\cite{Cuoco:2012yf,Harding:2012gk,Ajello:2011zi,Ajello:2013lka,Stecker:2010di} (possibly among other sources, including star-forming galaxies~\cite{Tamborra:2014xia,Ackermann:2012vca}, galaxy clusters~\cite{Zandanel:2014pva}, millisecond pulsars~\cite{Calore:2014oga,Hooper:2013nhl}, propagating ultra-high energy cosmic rays~\cite{Taylor:2015rla,Ahlers:2011sd}, and/or annihilating dark matter particles~\cite{Ackermann:2015tah,DiMauro:2015tfa,Ajello:2015mfa,Cholis:2013ena})\footnote{Previous work has shown that a large fraction of the total extragalactic gamma-ray background is dominated by emission from blazars, in particular at energies above $\sim$$50$ GeV~\cite{TheFermi-LAT:2015ykq}. We emphasize that the results of Ref.~\cite{Hooper:2016gjy} are not in conflict with this finding, as the IGRB makes up only about half of the total extragalactic background at these energies~\cite{Ackermann:2014usa}.}. This result was made possible by utilizing an empirical correlation that had been previously identified between the radio and gamma-ray luminosities of this class of sources~\cite{Inoue:2011bm,DiMauro:2013xta}. More quantitatively, Ref.~\cite{Hooper:2016gjy} concluded that unresolved radio galaxies account for $83.3^{+27.4}_{-10.1}\%$ of the $E_{\gamma} > 1$ GeV photons that make up Fermi's IGRB. This result is consistent with the findings of other recent work~\cite{Fornasa:2015qua,DiMauro:2016cbj,DiMauro:2015tfa,Ajello:2015mfa,Cholis:2013ena,Cavadini:2011ig,Siegal-Gaskins:2013tga}, including analyses based on cross-correlations of the IGRB with multi-wavelength data~\cite{Xia:2015wka,Cuoco:2015rfa,Shirasaki:2014noa,Shirasaki:2015nqp}.

The realization that radio galaxies dominate the IGRB has important implications for IceCube and their observed flux of high-energy astrophysical neutrinos.  In this paper, we demonstrate that if the gamma-ray emission observed from radio galaxies is generated through the interactions of cosmic-ray protons with gas, then one should expect these sources to also produce a spectrum of neutrinos that is qualitatively similar to that observed by IceCube. Given the large fraction of the IGRB that originates directly from these sources, we argue that any diffuse contribution from electromagnetic cascades must be suppressed, for example by very-high energy photon scattering taking place within or near the radio galaxies themselves, or through non-negligible synchrotron losses. Although scenarios in which IceCube's neutrinos are produced within the jets or lobes of active galactic nuclei (AGN) are possible, we instead consider a simple model in which high-energy cosmic rays are confined within the volumes of radio galaxies, where they interact with gas to generate the observed neutrino and gamma-ray fluxes (similar to earlier work within the context of starburst galaxies~\cite{Loeb:2006tw} and galaxy clusters~\cite{Berezinsky:1996wx}). This model predicts a cut-off in the neutrino spectrum at energies above approximately $E_{\nu}$~$\sim$~$1$-$100\, {\rm PeV}$, resulting from the transition between Kolmogorov diffusion and effective free-streaming. If we extrapolate the spectrum of cosmic rays that are accelerated by radio galaxies from $\sim$$10^8$ GeV to $\sim$$10^{11}$ GeV, we find that these sources can also generate the observed flux and spectrum of the ultra-high energy cosmic rays. It is possible that cosmic rays and/or neutrinos could be detected from individual radio galaxies in the future, making the most nearby and luminous examples of such sources (including Centaurus A, Centaurus B, and M 87) particularly promising targets of study.

\section{Gamma Rays and Neutrinos From Radio Galaxies}
\label{main}


The sum of the emission from all unresolved radio galaxies leads to a gamma-ray flux that is given by:
\begin{eqnarray}
\label{totspec}
\frac{dF_{\gamma}}{dE_{\gamma} \, d\Omega} =   \int dz \frac{d^2V}{dz \, d\Omega} \int \frac{dF_\gamma}{dE_{\gamma}}  \frac{dL_\gamma}{L_\gamma \log (10)} \rho_\gamma (L_\gamma,z)(1-\omega (F_\gamma (L_\gamma ,z)))\exp (-\tau_{\gamma}(E_{\gamma},z)), \nonumber
\end{eqnarray}
where $d^2V/dz \, d\Omega$ is the co-moving volume element, and $dF_{\gamma}/dE_{\gamma}$ is the spectrum of gamma-rays from a radio galaxy of luminosity $L_{\gamma}$ and located at redshift $z$. The function $\omega$ represents Fermi's point source detection efficiency~\cite{Collaboration:2010gqa}, which accounts for the fact that resolved radio galaxies do not contribute to the diffuse gamma-ray background\footnote{In the case of radio galaxies, this calculation depends very little on the precise form of $\omega$, as only a very small fraction of the total gamma-ray flux originates from above or near threshold sources.}. The attenuation of the gamma-ray spectrum from scattering with the extragalactic background light is characterized by the optical depth, $\tau_{\gamma} (E_{\gamma}, z)$, for which we adopt the model of Ref.~\cite{2010ApJ...712..238F}.

In a recent study~\cite{Hooper:2016gjy}, we refined the empirical correlation between the radio and gamma-ray emission detected from radio galaxies (see also Refs.~\cite{Inoue:2011bm,DiMauro:2013xta}) and combined this information with the measured radio luminosity function and redshift distribution~\cite{Willott:2000dh} to determine the gamma-ray luminosity function of radio galaxies, $\rho_{\gamma} (L_\gamma,z)$, and ultimately the total contribution from all unresolved radio galaxies to the diffuse gamma-ray background. In doing so, we found that this class of sources dominates the unresolved extragalactic gamma-ray flux, accounting for $83.3^{+27.4}_{-10.1}\%$ of the isotropic gamma-ray background (IGRB) observed by Fermi~\cite{Ackermann:2014usa} above 1 GeV. 

At this time, it is not entirely clear whether the gamma-ray emission observed from radio galaxies results from hadronic (pion production) or leptonic (inverse Compton) processes. If hadronic processes are responsible, however, then high-energy neutrinos will accompany the observed gamma-rays.  In this section, we will focus on models in which gamma rays and neutrinos are generated through the interactions of cosmic-ray protons with gas, and the subsequent decays of charged and neutral pions ($\pi^{+} \rightarrow e^{+} \nu_e \nu_{\mu} \bar{\nu}_{\mu}$, $\pi^{-} \rightarrow e^{-} \bar{\nu}_e \nu_{\mu} \bar{\nu}_{\mu}$, $\pi^0 \rightarrow \gamma \gamma$). The spectra of these gamma rays and neutrinos are generally predicted to have a common shape (prior to attenuation), and with relative fluxes given by $F_{\nu}/F_{\gamma} = 2 \times (3/4) = 3/2$, where the factors of 2 and 3/4 result from the ratio of charged-to-neutral pions that are produced in such interactions and from the fact that three of the four decay products of a charged pion are neutrinos.

\begin{figure}
\includegraphics[keepaspectratio,width=0.49\textwidth]{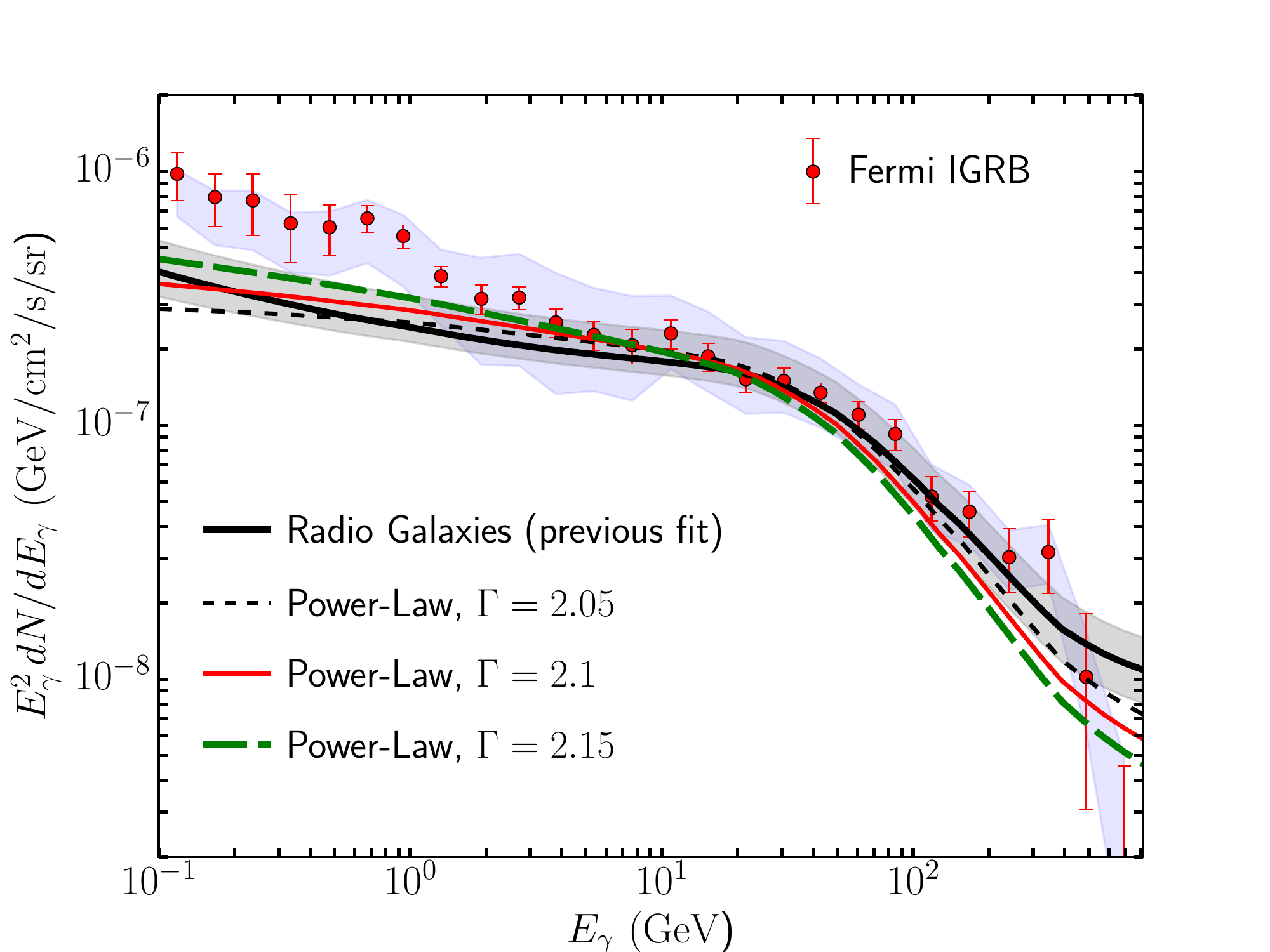}
\includegraphics[keepaspectratio,width=0.49\textwidth]{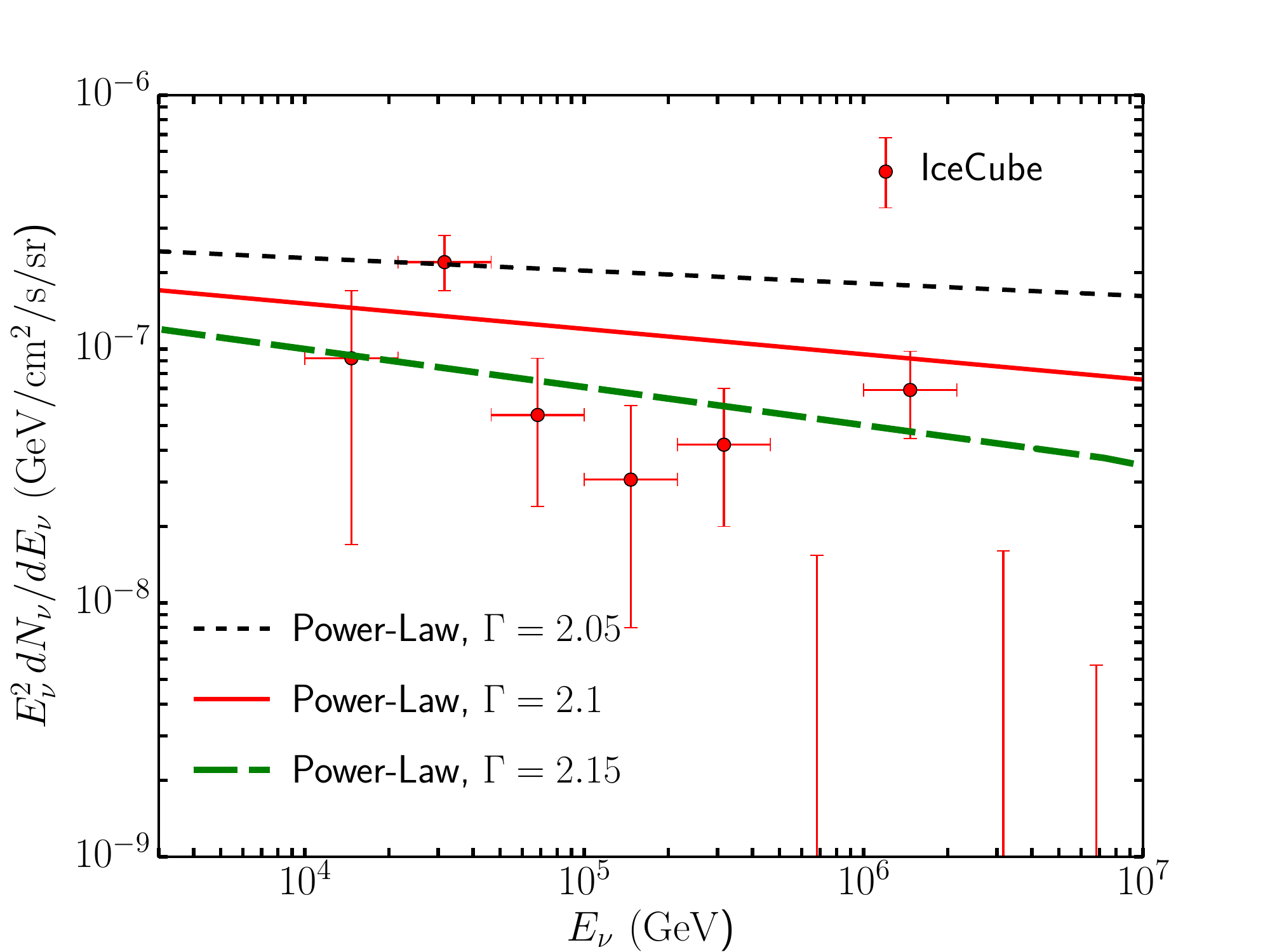}
\caption{Left Frame: The contribution to the diffuse gamma-ray background from unresolved radio galaxies, as determined previously~\cite{Hooper:2016gjy} (solid black and surrounding grey band), and compared to power-law spectra (prior to attenuation) with three values of the spectral index (dashed black, solid red, long-dashed green). We also show the measurement of the isotropic gamma-ray background (IGRB) as reported by the Fermi Collaboration~\cite{Ackermann:2014usa}. The error bars in this figure include both the statistical uncertainty and the systematic uncertainties associated with the effective area and cosmic ray background subtraction, while the light blue shaded band reflects the systematic uncertainties associated with the modeling of the Galactic foreground emission. Right Frame: The (all-flavor) neutrino spectrum from the same radio galaxy models shown in the left frame, assuming that 1) the gamma-ray emission is generated through proton-proton collisions, and 2) the spectrum of this emission can be extrapolated to the energy range measured by IceCube. The fact that these extrapolated models are able to approximately accommodate the diffuse neutrino flux reported by the IceCube Collaboration~\cite{Aartsen:2015knd} is suggestive of a scenario in which high-energy protons in radio galaxies are responsible for both the majority of the observed isotropic gamma-ray background and the diffuse high-energy neutrino flux.}
\label{spec1}
\end{figure}

In the left frame of Fig.~\ref{spec1}, we show the contribution to the diffuse gamma-ray background from unresolved radio galaxies as determined in Ref.~\cite{Hooper:2016gjy} and compare this to Fermi's measurement of the IGRB~\cite{Ackermann:2014usa}. In Ref.~\cite{Hooper:2016gjy}, the total gamma-ray spectrum from radio galaxies (prior to attenuation) was calculated as a weighted sum of power-laws. For simplicity (and to facilitate a more straightforward extrapolation), we will instead consider spectra that are described by a single power-law. In the left frame of Fig.~\ref{spec1}, we find that a power-law index of $\Gamma \simeq 2.1$ provides a reasonable match to that obtained in Ref.~\cite{Hooper:2016gjy}, especially in the energy range best measured by Fermi ($\sim$$0.7$-$10$ GeV). In each case, we have accounted for gamma-ray attenuation using the infrared background model described in Ref.~\cite{2010ApJ...712..238F}. 

In the right frame of Fig.~\ref{spec1}, we show the (all-flavor) neutrino spectrum for the same three power-law models, normalized as in the left frame, and assuming that the power-law spectra extend to the energy range measured by IceCube. When these extrapolated spectra are compared to the results reported by the IceCube Collaboration~\cite{Aartsen:2015knd} (see also Refs.~\cite{Aartsen:2015rwa,Aartsen:2014gkd,Aartsen:2013jdh,Aartsen:2013bka}), we find reasonable agreement. This result is suggestive, and provides support for scenarios in which high-energy protons in active galaxies are responsible for both Fermi's observed isotropic gamma-ray background and IceCube's high-energy neutrino flux.

\begin{figure}
\includegraphics[keepaspectratio,width=0.49\textwidth]{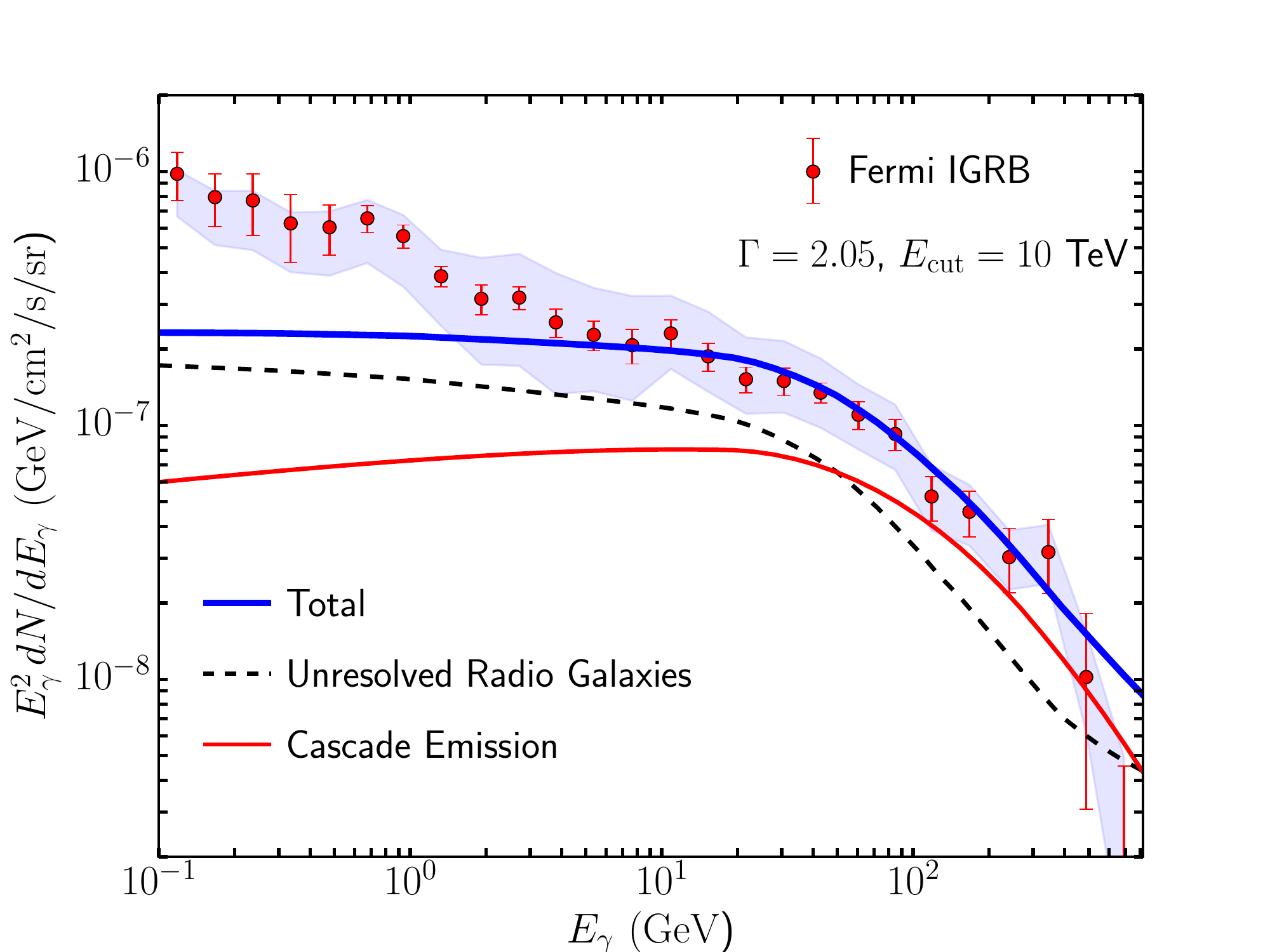}
\includegraphics[keepaspectratio,width=0.49\textwidth]{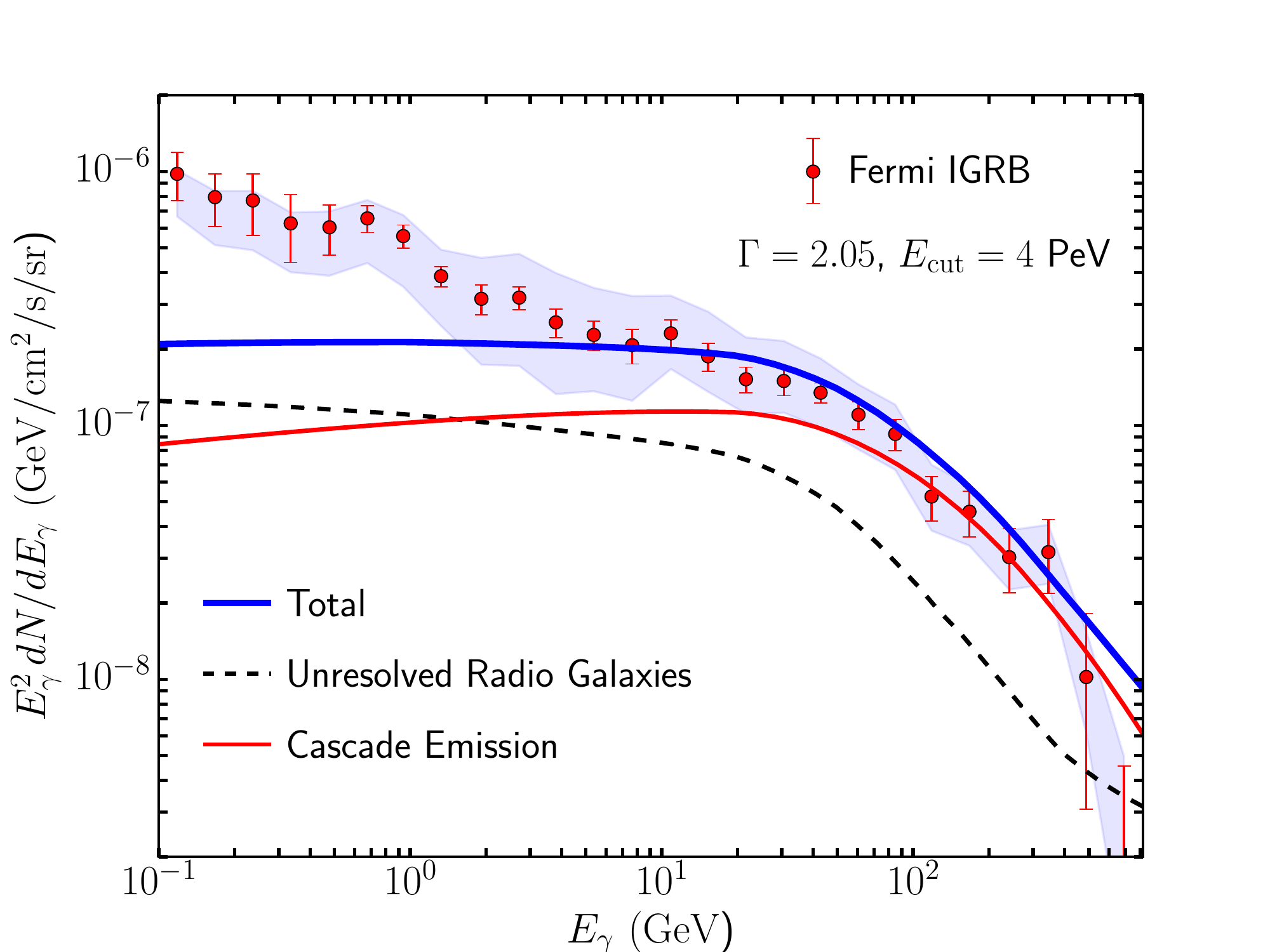}\\
\includegraphics[keepaspectratio,width=0.49\textwidth]{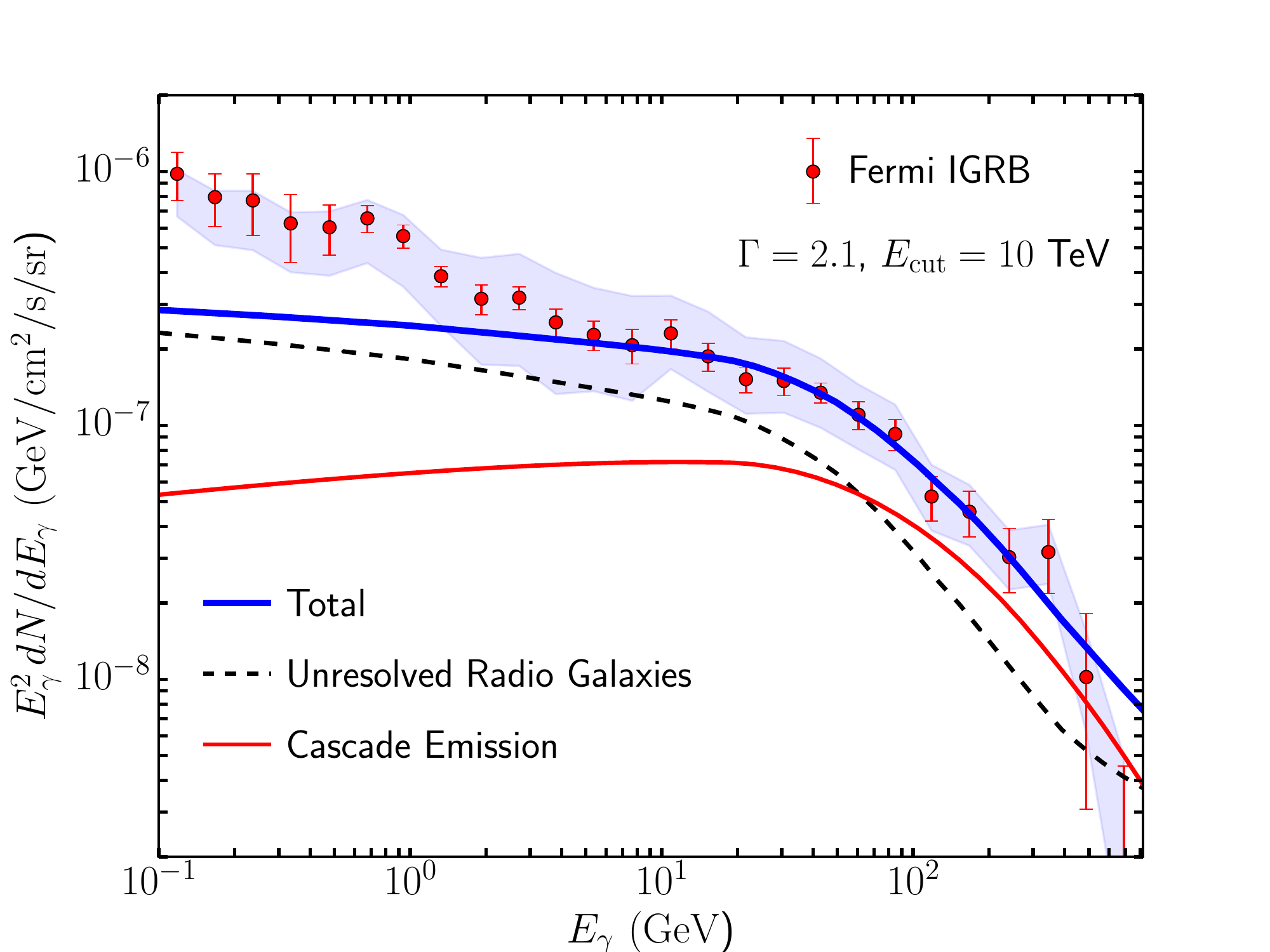}
\includegraphics[keepaspectratio,width=0.49\textwidth]{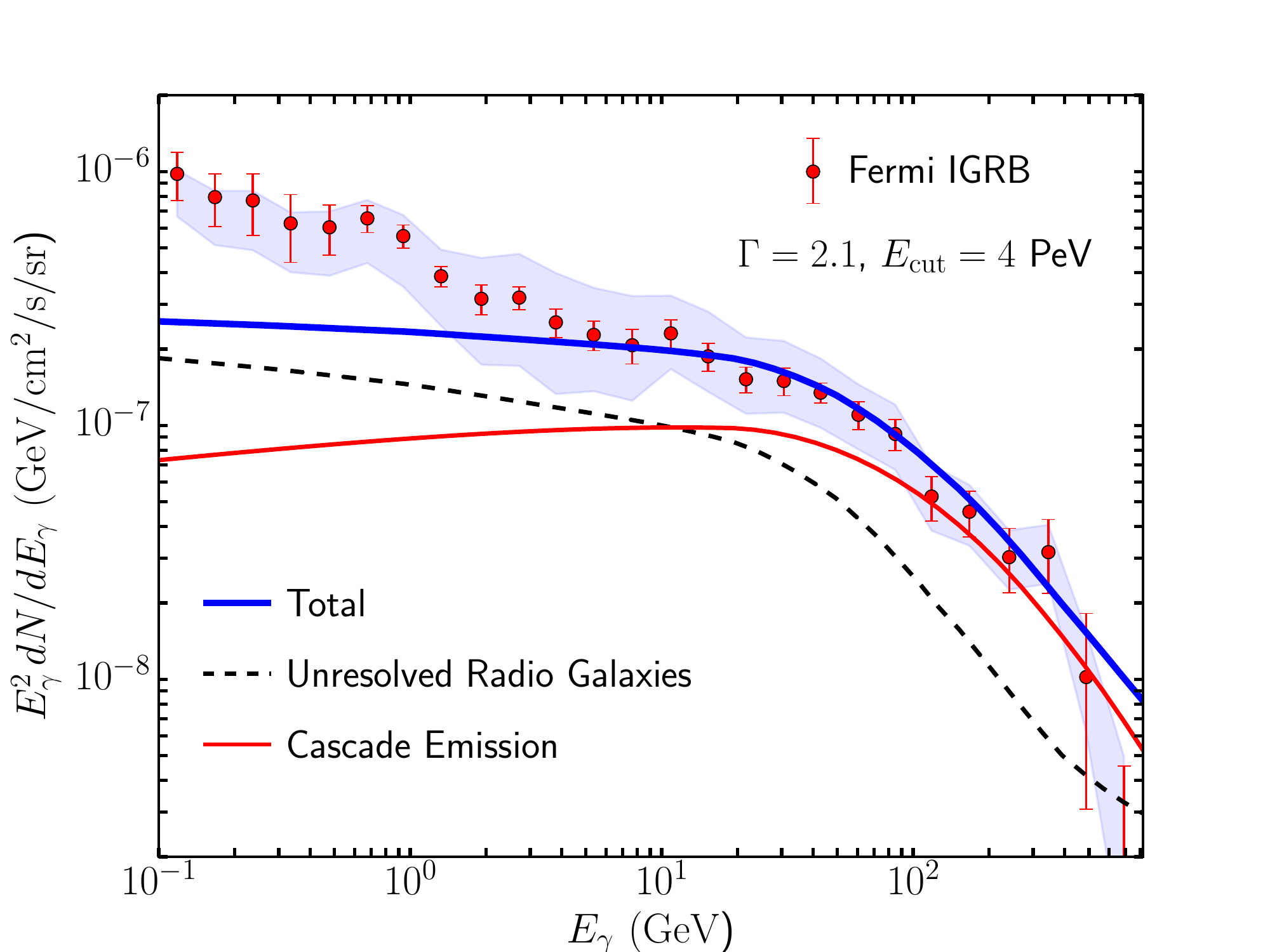}\\
\includegraphics[keepaspectratio,width=0.49\textwidth]{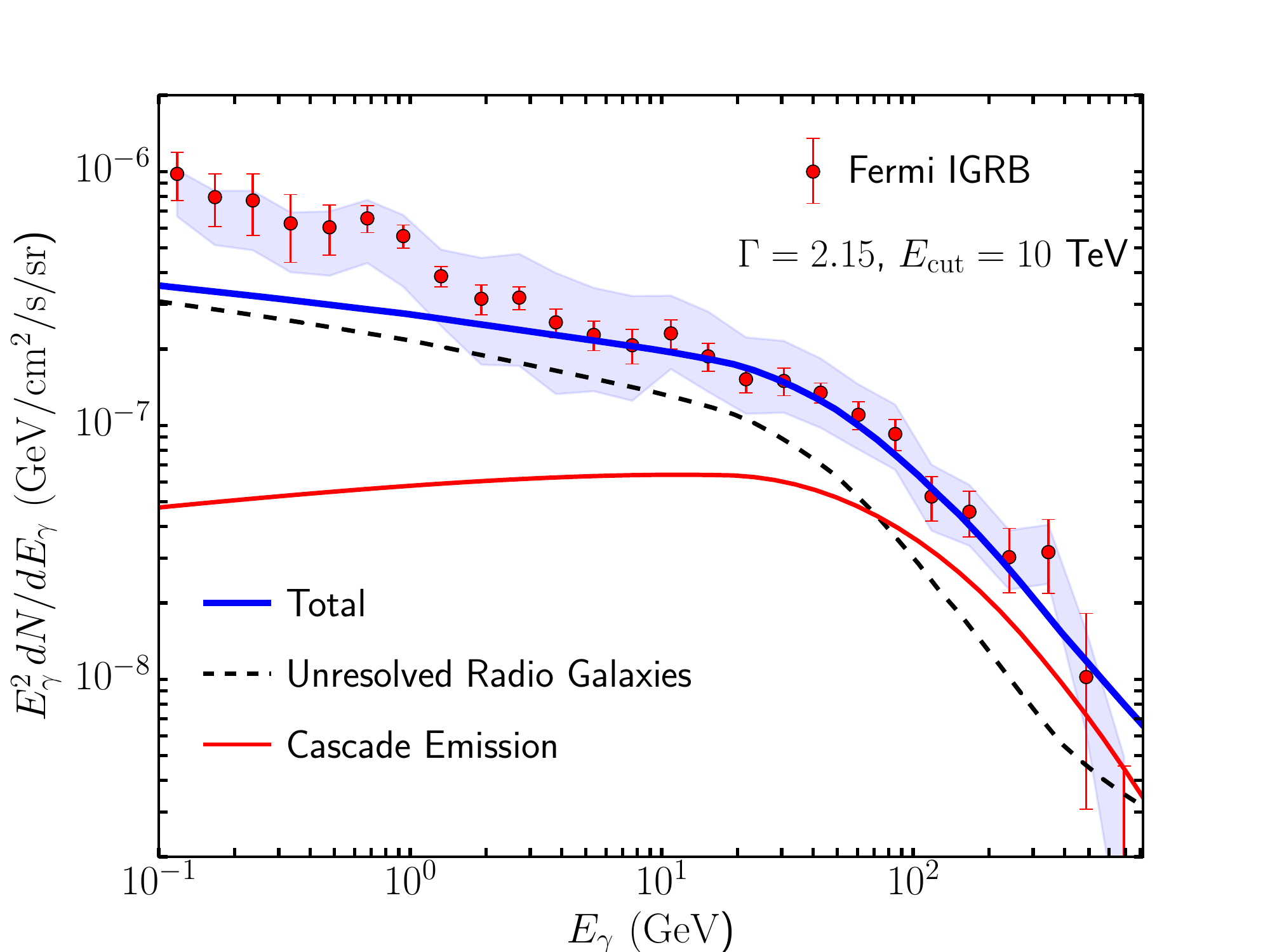}
\includegraphics[keepaspectratio,width=0.49\textwidth]{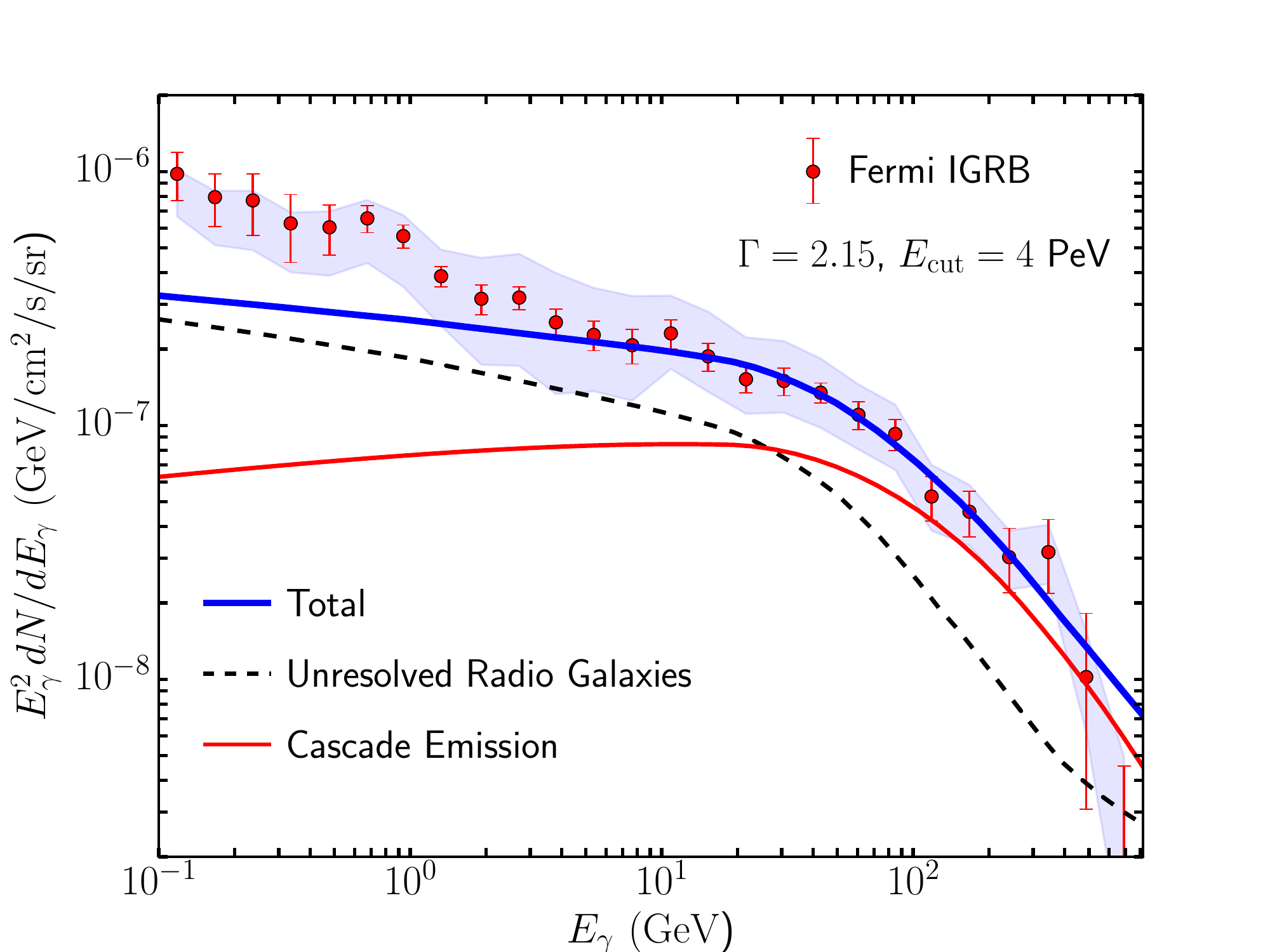}
\caption{The contribution to the diffuse gamma-ray background from unresolved radio galaxies, and from the emission generated in the electromagnetic cascades initiated by very-high energy photons. We show results for three choices of the spectral index, and for two choices of the cut-off energy in the initial (unattenuated) gamma-ray spectrum, $E_{\rm cut}$. The total flux in each frame is normalized to the measured intensity of the IGRB. Here, we assume that 100\% of the total energy in electromagnetic cascades goes into the production of diffuse gamma rays ($f_{\rm cas}=1$).}
\label{withcas}
\end{figure}

Radio galaxies not only produce gamma rays directly, but also through the electromagnetic cascades that result from the scattering of very-high energy photons with radiation. The intensity of any diffuse cascade emission is significantly limited by the results of Ref.~\cite{Hooper:2016gjy}, however, which allow for only a relatively small fraction of the IGRB to originate from cosmologically induced cascades. The predicted intensity of the cascade contribution depends on the redshift distribution of the very-high energy gamma-ray sources, and on the maximum energy to which their (unattenuated) spectrum extends. If the neutrino and gamma-ray spectra injected from radio galaxies does in fact extend up to $\sim$PeV energies or above, one might expect the resulting cascade emission to constitute a significant fraction of the IGRB~\cite{Murase:2013rfa,Chang:2016ljk,Murase:2015xka,Kalashev:2014vra}, in possible conflict with the findings of Ref.~\cite{Hooper:2016gjy}. 

\begin{figure}
\includegraphics[keepaspectratio,width=0.49\textwidth]{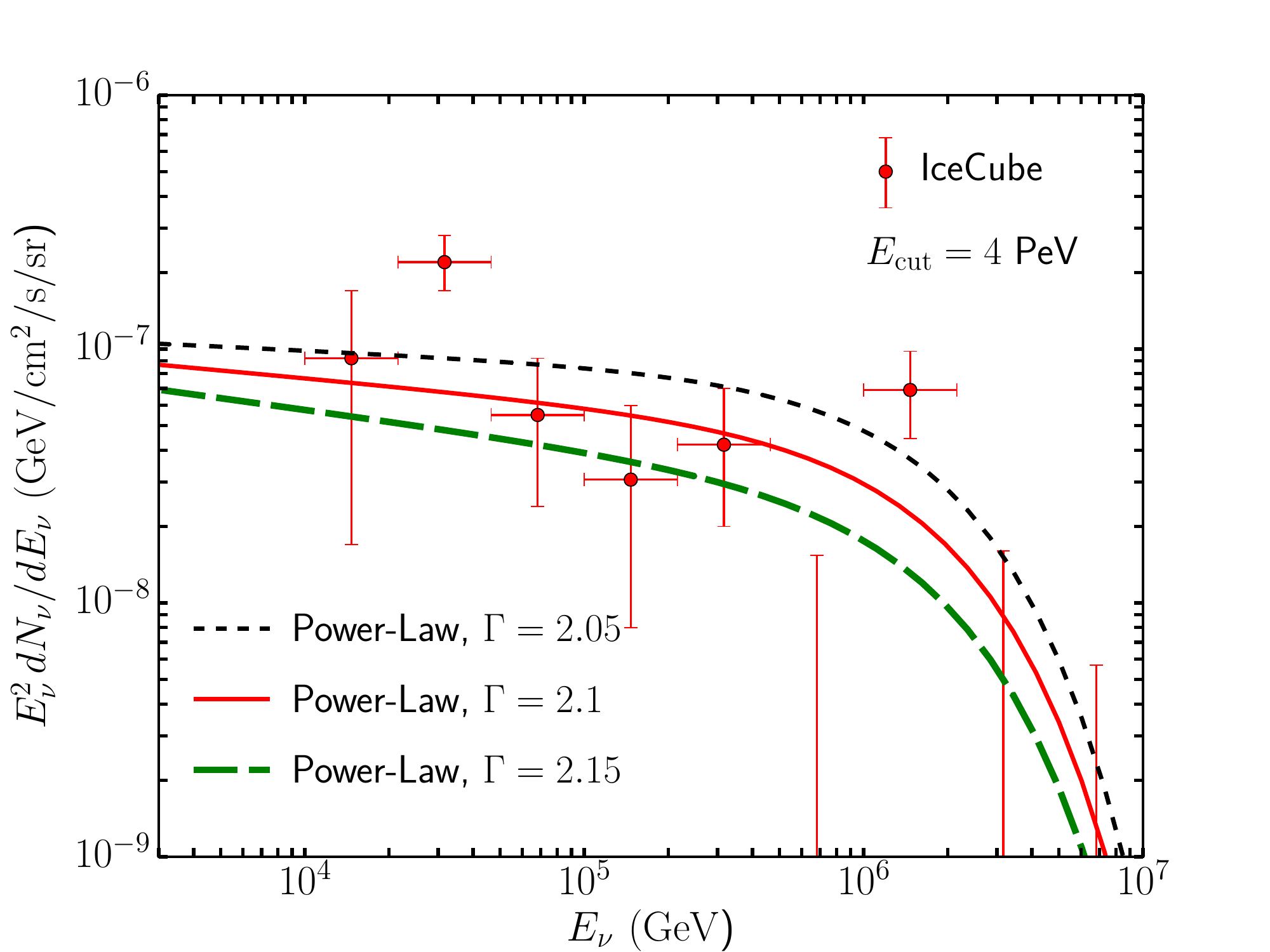}
\caption{The (all-flavor) neutrino spectra for the models shown in Fig.~\ref{withcas}, for the case of $E_{\rm cut}=4$ PeV and $f_{\rm cas}=1$.}
\label{spec3}
\end{figure}

In Fig.~\ref{withcas}, we plot the contributions to the IGRB from radio galaxies and from the corresponding cascade emission (for details regarding the cascade calculation, see Refs.~\cite{Murase:2011yw,Murase:2012xs,Murase:2011cy,Murase:2012df,Berezinsky:2016feh}), for three values of the spectral index and for two choices of the maximum gamma-ray energy, $E_{\rm cut}$. Even for the minimum value of $E_{\rm cut} = 10$ TeV,\footnote{As a number of radio galaxies have been observed by ground-based gamma-ray telescopes at energies of $\sim$10 TeV (see, for example, Refs.~\cite{Acciari:2008ah,Dyrda:2015hxa,Acciari:2009rs,Galante:2009ie,Aleksic:2013kaa,:2012uma}) we take this to be the minimum acceptable value of $E_{\rm cut}$.} the contribution from electromagnetic cascades is in some tension with Ref.~\cite{Hooper:2016gjy}, although concordance may be possible if $\Gamma \gsim 2.15$. This tension is made significantly worse, however, if the gamma-ray spectrum from these sources extends to $E_{\rm cut}\sim$~PeV, as would be required to accommodate IceCube's measured neutrino spectrum (see Fig.~\ref{spec3}).

\begin{figure}
\includegraphics[keepaspectratio,width=0.49\textwidth]{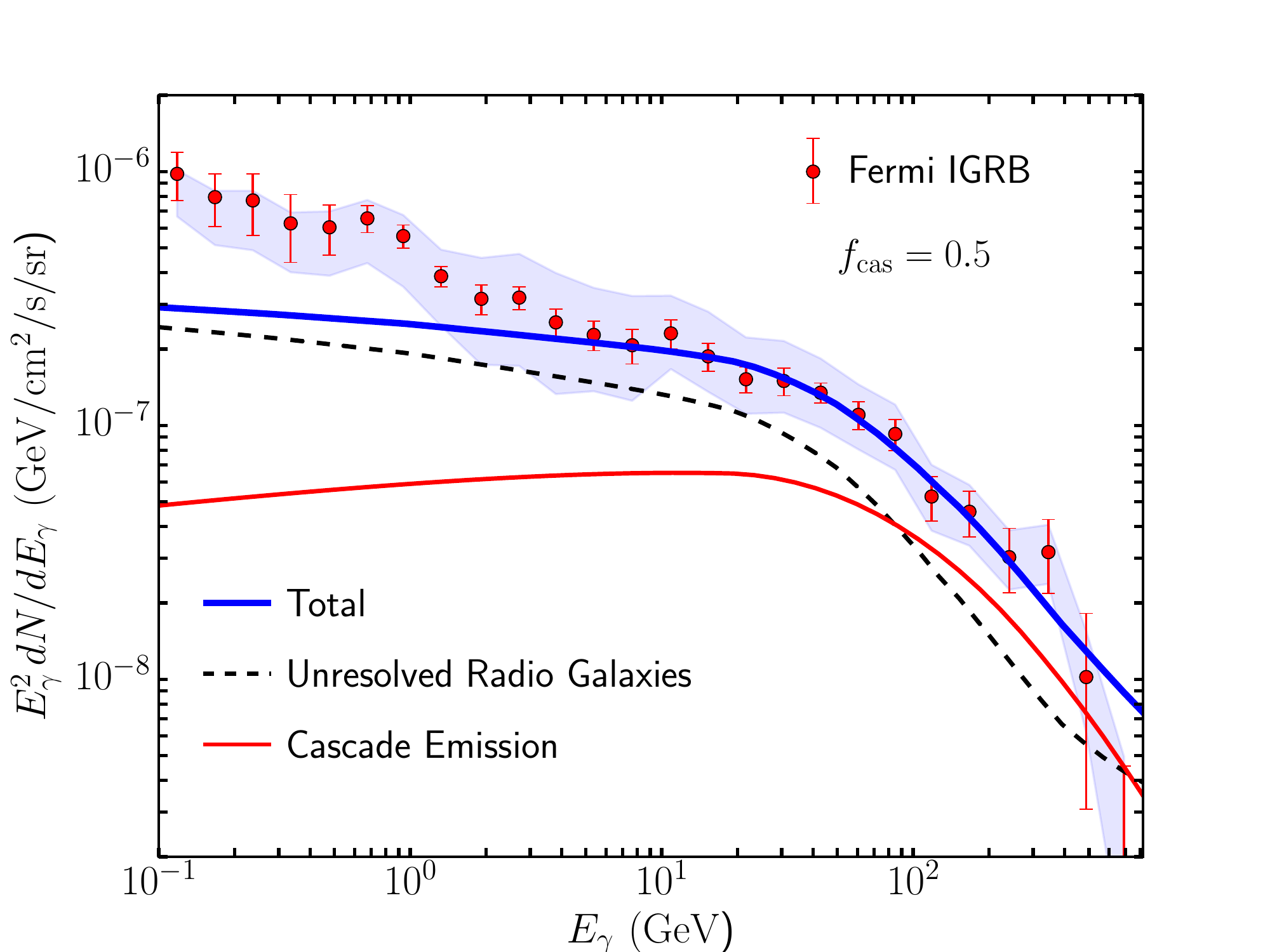}
\includegraphics[keepaspectratio,width=0.49\textwidth]{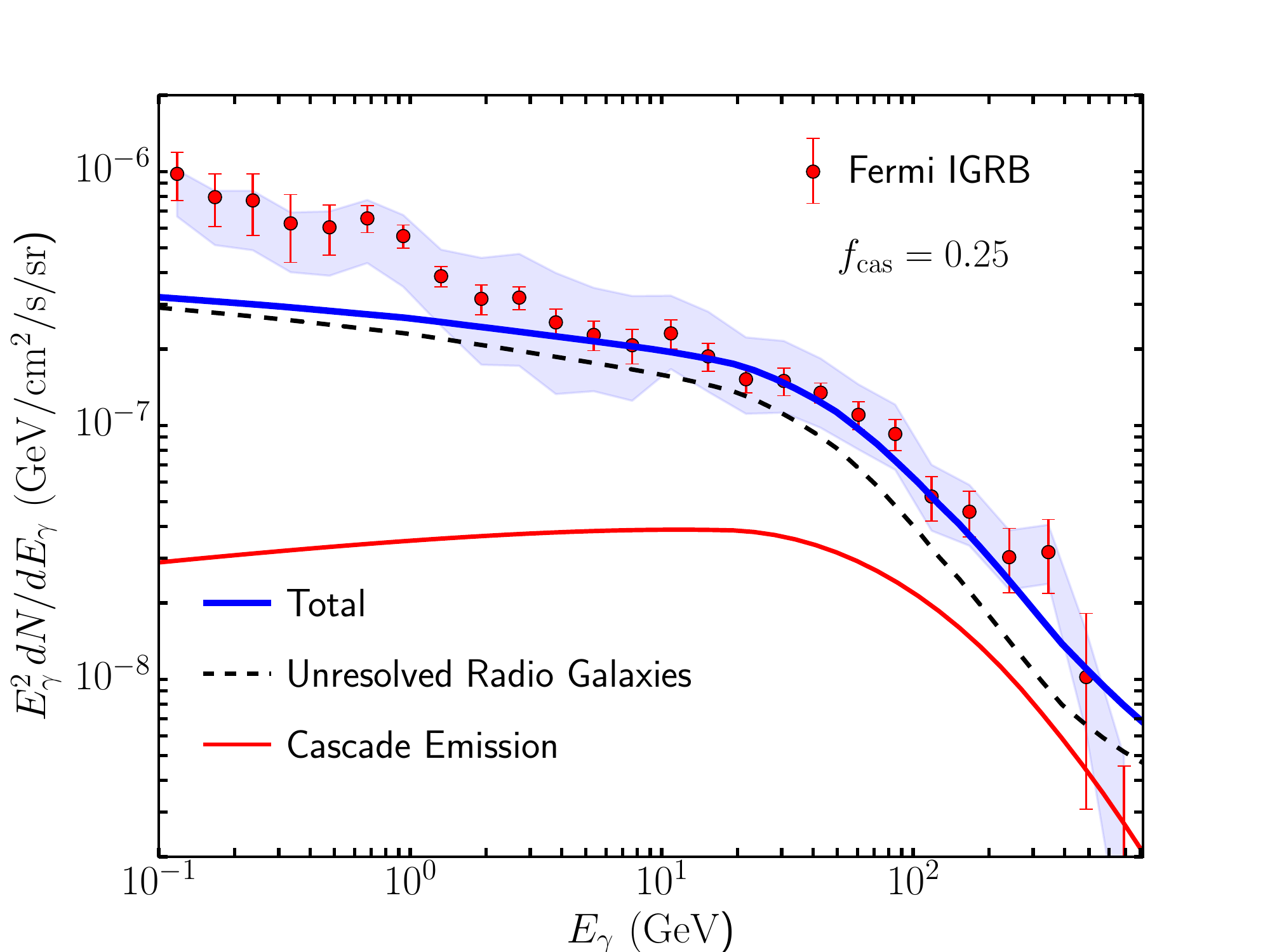}\\
\includegraphics[keepaspectratio,width=0.49\textwidth]{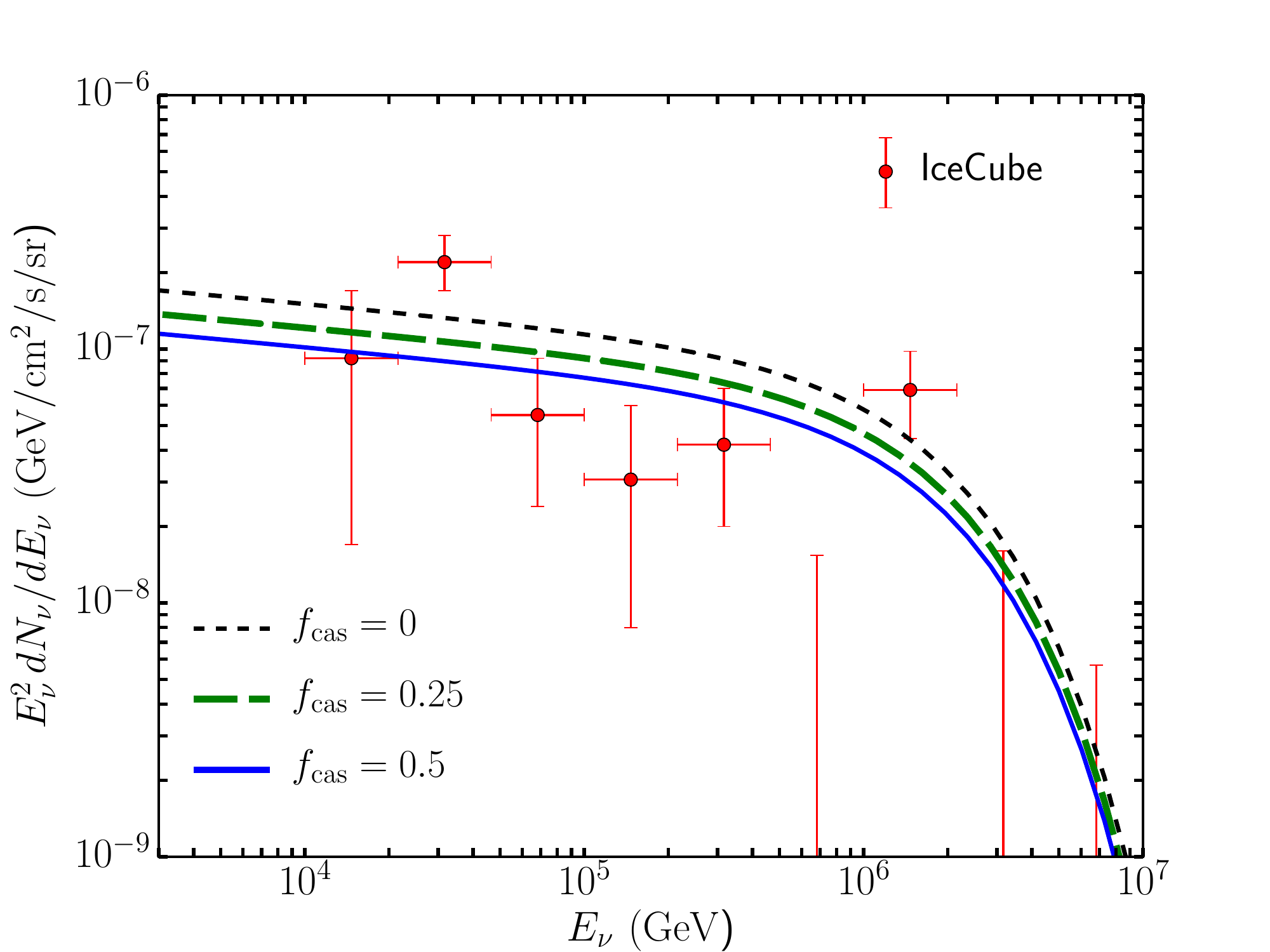}
\caption{Upper frames: As in Fig.~\ref{withcas}, but assuming that 50\% or 25\% of the total energy in electromagnetic cascades goes into diffuse gamma rays ($f_{\rm cas}=0.5$, 0.25). In each case, we have taken $\Gamma=2.1$ and $E_{\rm cut}=4$ PeV. Lower Frame: The (all-flavor) neutrino spectra for the same range of models, with three values of $f_{\rm cas}$. Fermi's IGRB (including the fraction that originates directly from radio galaxies~\cite{Hooper:2016gjy}) and IceCube's diffuse neutrino spectrum can be simultaneously accommodated in this class of scenarios.}
\label{spec4}
\end{figure}

This tension can be relieved, however, if a non-neglibile fraction of the high photons initiating electromagnetic cascades do so within or nearby the location of their parent radio galaxy. If this is the case, much of the cascade emission will point in the direction of the source radio galaxy, and will thus simply be included as part of the emission that we label in Fig.~\ref{withcas} as being from ``unresolved radio galaxies'' (as opposed to that from ``cascade emission''). Photons with an energy of $\sim$1 PeV or more are predicted to scatter with radiation before escaping a galaxy, even in the case of Milky Way-like systems~\cite{Zhang:2005tp,Moskalenko:2005ng,Gupta:2011aw}. In active galaxies, with higher densities of infrared radiation, such interactions will be more efficient, plausibly leading to the scattering of most of the gamma-rays with $E_{\gamma} \gsim 10$ TeV. We also mention that synchrotron cooling could reduce the energy in some electromagnetic cascades, suppressing the total amount of energy in diffuse gamma-rays (for a discussion of synchrotron and inverse Compton energy losses, see Ref.~\cite{Lee:1996fp}).


With this in mind, we plot in the upper two frames of Fig.~\ref{spec4} the gamma-ray spectrum from unresolved radio galaxies and from their corresponding cascade emission, in scenarios in which 50\% or 25\% of the total cascade energy goes into diffuse gamma rays (those not directed at their source). In both cases, we have taken $E_{\rm cut}=$4 PeV, enabling these sources to generate the neutrino spectrum observed by IceCube (as shown in the lower frame of this figure).  From this comparison, we conclude that Fermi's IGRB and IceCube's diffuse neutrino spectrum can be simultaneously generated by radio galaxies in scenarios in which a significant fraction of the electromagnetic cascades are initiated in or nearby their source galaxy ($f_{\rm cas} \sim 0.1-0.5$).


\section{Toward a Physical Model for Cosmic Ray Diffusion and Scattering in Radio Galaxies}
\label{toward}

Active galaxies have long been considered to be one of the most promising classes of sources for the highest energy cosmic rays, with magnetic fields and geometries that make them potentially capable of accelerating protons and/or nuclei up to the highest observed energies~\cite{Hillas:1985is}. While within the jet of an active galactic nuclei (AGN), cosmic rays may or may not efficiently scatter with radiation fields, depending on the spectrum and energy density of the target photons (for additional discussion, see Ref.~\cite{Murase:2015ndr}). Interactions between cosmic rays and gas are generally not expected to be important within AGN jets (for exceptions, see the models described in Refs.~\cite{Tjus:2014dna,Kimura:2014jba}). Thus, as long as the radiation fields are not overly dense, we expect most of the cosmic rays accelerated in these environments to escape into the surrounding galaxy.



After leaving the jet, high-energy protons will diffuse through the volume of their parent galaxy (or galaxy cluster), moving under the influence of the magnetic fields in a way that resembles a random walk.  Over a time, $t$, a typical particle will be displaced by a distance of $d_{\rm dif} \sim 2\sqrt{D(E_p) t}$, where $D(E_p)$ is the energy dependent diffusion coefficient. Adopting a Kolmogorov spectrum of magnetic inhomogeneities, the diffusion coefficient takes the following form:
\begin{eqnarray}
\label{dif}
D(E_p) &=& \frac{1}{3} c \, l_c \bigg(\frac{r_L}{l_c}\bigg)^{1/3}  \\
& \approx & 1.5 \times 10^{30} \, {\rm cm}^2/{\rm s} \, \bigg(\frac{E_p}{{\rm PeV}}\bigg)^{1/3} \bigg(\frac{l_c}{{\rm kpc}}\bigg)^{2/3} \bigg(\frac{ \mu{\rm G}}{B}\bigg)^{1/3},  \nonumber 
\end{eqnarray}
where $l_c$ is the coherence length of the magnetic field and $r_L$ is the Larmor radius of the propagating cosmic ray. We note that for $l_c^2/B \simeq {\rm kpc}^2/\mu{\rm G}$, this diffusion coefficient matches the value measured for GeV-TeV cosmic rays in the Milky Way~\cite{Simet:2009ne,Trotta:2010mx}. The expression given in Eq.~\ref{dif} is expected to be valid only in the limit of $r_L \ll l_c$, corresponding to $E_p \ll 0.9\, {\rm EeV} \times (l_c/{\rm kpc}) (B/\mu{\rm G})$~\cite{Aloisio:2004jda}. At energies around or above this value, the diffusion coefficient increases rapidly, enabling cosmic rays to effectively free-stream out of their parent galaxy. We thus anticipate that the spectrum of cosmic rays confined within such a galaxy will undergo a sharp cutoff at an energy on the order of $E^{\rm max}_p \sim 0.1-1\, {\rm EeV} \times (l_c/{\rm kpc}) (B/\mu{\rm G})$.

%

From the above diffusion coefficient, we can estimate the length of time that a typical cosmic ray-proton will remain confined within the volume of its parent galaxy:
\begin{eqnarray}
t_{\rm esc} &\sim& \frac{d^2_{\rm dif}}{2 D(E_p)} \\
& \sim & 7.9 \times 10^{13} \, {\rm s} \times \bigg(\frac{d_{\rm dif}}{10\,{\rm kpc}}\bigg)^2 \bigg(\frac{{\rm PeV}}{E_p}\bigg)^{1/3}  \bigg(\frac{{\rm kpc}}{l_c}\bigg)^{2/3} \bigg(\frac{B}{\mu{\rm G}}\bigg)^{1/3}. \nonumber 
\end{eqnarray}

Over this period of time, the probability that a given proton will scatter with the ambient gas is $P(E_p) = 1-e^{-\tau_{pp}(E_p)}$, where the optical depth is given by:
\begin{eqnarray}
\label{tau}
\tau_{pp}(E_p) &=& \sigma_{pp}(E_p) \, c \, t_{\rm esc} \, n_{\rm gas} \\
& \approx & 0.054 \times \bigg(\frac{n_{\rm gas}}{0.3 \, {\rm cm}^{-3}}\bigg) \bigg(\frac{d_{\rm dif}}{10\,{\rm kpc}}\bigg)^2 \bigg(\frac{{\rm PeV}}{E_p}\bigg)^{1/3}  \bigg(\frac{{\rm kpc}}{l_c}\bigg)^{2/3} \bigg(\frac{B}{\mu{\rm G}}\bigg)^{1/3}. \nonumber
\end{eqnarray}
Here, $n_{\rm gas}$ represents the average number density of target nucleons within the volume of the diffusion region. From this equation, we learn that for $E_p \gsim 10^2$ GeV, $\tau$ is less than one, and the probability of scattering can be approximated by $P(E_p) \simeq \tau_{pp}(E_p) \propto E_p^{-1/3}$.


In the energy range of interest, the average number of pions produced in a proton-proton collision scales as 
$N_{\pi} \propto E_p^{1/4}$, while the average fraction of energy carried by a given pion scales as $\langle E_{\pi} \rangle /E_p \propto  E_p^{-1/4}$~\cite{1994A&A...286..983M,Tjus:2014dna}.\footnote{For useful parameterizations of the neutrino and gamma-ray spectra from proton-proton collisions, based on fits to accelerator data, we direct the reader to Ref.~\cite{Kamae:2006bf}.} The neutrinos and gamma rays generated in the decays of these pions are further reduced in energy by factors of 4 and 2, respectively, following from the number of particles in their decays. If we consider a power-law spectrum of protons, $dN_p/dE_p =A_p E_p^{-\Gamma_p}$, it follows that the resulting neutrinos and gamma-rays will take on power-law spectra with an index of $\Gamma_{\nu,\gamma} = -(4/3)\Gamma_p + (2/3)$~\cite{Tjus:2014dna}.

Together, this leads to the following spectra for neutrinos and gamma rays:
\begin{eqnarray}
\label{gammaspec}
\frac{dN_{\gamma}}{dE_{\gamma}} &\approx& A_{\gamma} \times \bigg(\frac{n_{\rm gas}}{0.3 \, {\rm cm}^{-3}}\bigg) \bigg(\frac{d_{\rm dif}}{10\,{\rm kpc}}\bigg)^2 \bigg(\frac{{\rm kpc}}{l_c}\bigg)^{2/3} \bigg(\frac{B}{\mu{\rm G}}\bigg)^{1/3}  \bigg(\frac{E_{\gamma}}{{\rm GeV}}\bigg)^{-\frac{4}{3}\Gamma_p + \frac{1}{3}},
\end{eqnarray}
and
\begin{eqnarray}
\frac{dN_{\nu}}{dE_{\nu}} &\approx& A_{\gamma} \times \frac{3}{2} \times \bigg(\frac{n_{\rm gas}}{0.3 \, {\rm cm}^{-3}}\bigg) \bigg(\frac{d_{\rm dif}}{10\,{\rm kpc}}\bigg)^2 \bigg(\frac{{\rm kpc}}{l_c}\bigg)^{2/3} \bigg(\frac{B}{\mu{\rm G}}\bigg)^{1/3}  \bigg(\frac{E_{\nu}}{{\rm GeV}}\bigg)^{-\frac{4}{3}\Gamma_p + \frac{1}{3}}, \nonumber
\end{eqnarray}
where $A_{\gamma}$ is related to $A_p$ by $\int A_p E_p^{-\Gamma_p+1} \tau_{pp}(E_p) dE_p = (5/3) \int A_{\gamma} E_{\gamma}^{(-4/3)\Gamma_p+(4/3)}  dE_{\gamma}$. These power-laws are expected to extend up to energies of $E_{\gamma} \sim (8-80) \, {\rm PeV} \times (l_c/{\rm kpc}) (\mu{\rm G}/B)$ and $E_{\nu} \sim (4-40) \, {\rm PeV} \times (l_c/{\rm kpc}) (\mu{\rm G}/B)$, above which $r_L \gsim l_c$ and protons diffuse with $D \propto E_p^2$, leading to a steepening of the resulting gamma-ray and neutrino spectra by an additional power of $E_{\gamma, \nu}^{-5/3}$~\cite{Aloisio:2004jda}. We note that the spectral cut-off predicted in this scenario is distinct from that found for the model described in Ref.~\cite{Tjus:2014dna}, where it was argued that the detection of such a feature would disfavor radio galaxies as the sources of IceCube's neutrinos.

To accommodate the required spectral index for gamma rays and neutrinos, $\Gamma_{\gamma} \simeq 2.1$ (see Sec.~\ref{main}), Eq.~\ref{gammaspec} indicates that protons must be injected into their parent galaxies with an index of $\Gamma_p \simeq 1.8$. This value is within the range favored to explain the observed ultra-high energy cosmic ray spectrum, as is the measured redshift distribution of radio galaxies~\cite{Taylor:2015rla,Taylor:2013gga,Taylor:2011ta}. 

In light of these features, radio galaxies appear to be excellent candidates for the sources of the ultra-high energy cosmic rays. The famous calculation by Waxman and Bahcall can be used to relate the fluxes of ultra-high energy cosmic rays and neutrinos (for the case of proton-proton collisions and an $E^{-2}$ spectral shape)~\cite{Bahcall:1999yr,Waxman:1998yy}:
\begin{eqnarray}
[E_{\nu}^2 \Phi_\nu ]_{\rm WB} &\approx& \xi_Z \, \tau_{pp} \, t_H \frac{c}{8\pi} \, E^2_{\rm CR} \frac{d\dot{N}_{\rm CR}}{d\dot{E}_{\rm CR}} \nonumber \\
&\approx& 8.0 \times 10^{-8} \, {\rm GeV}\, {\rm cm}^{-2}\,  {\rm s}^{-1}  \, {\rm sr}^{-1}  \times \bigg(\frac{\xi_Z}{4.8}\bigg) \, \bigg(\frac{\tau_{pp}}{0.054}\bigg).
\end{eqnarray}
In this expression, $\xi_Z$ is a factor which accounts for redshift dependent source evolution. For the observed redshift distribution of FR-I type radio galaxies~\cite{Willott:2000dh}, we calculate $\xi_Z \simeq 4.8$. $t_H$ is the Hubble time and $E^2_{\rm CR} d\dot{N}_{\rm CR}/d\dot{E}_{\rm CR} \approx 10^{44}$ erg Mpc$^{-3}$ yr$^{-1}$ is the (local) injection rate of ultra-high energy ($>10^{19}$ eV) cosmic ray sources. For the optical depth given in Eq.~\ref{tau}, we find excellent agreement with IceCube's measured flux. In other words, if we were to simply extrapolate the cosmic ray spectrum from radio galaxies from energies of $\sim$$10^{17}$ eV (as required to generate IceCube's observed flux) to $\sim$$10^{20}$ eV, this spectrum would provide a reasonable match to that of the ultra-high energy cosmic rays.

\section{Discussion and Conclusions}
\label{discussion}

At this point in time, it has become possible to make some rather far-reaching and model-independent statements regarding the origin of IceCube's neutrino flux. Of particular importance is Fermi's measurement of the isotropic gamma-ray background (IGRB), which significantly restricts the range of scenarios that could potentially be responsible for the observed neutrinos. For a wide range of spectral shapes, source distributions, and interactions ($\gamma p$, $pp$), it has been shown that models capable of generating the spectrum measured by IceCube also generate a diffuse flux of gamma rays that approximately saturates or exceeds that observed by Fermi. And although this conclusion can be mitigated, to some extent, by considering sources that are not entirely transparent to very-high energy gamma rays~\cite{Murase:2015xka} (as we did in Sec.~\ref{main}, by considering in-galaxy pair production), this argument appears to favor a common origin for both IceCube's neutrino flux and the majority of the IGRB. 

This connection is particularly powerful in light of the recent results of Ref.~\cite{Hooper:2016gjy}, which found that the IGRB is dominated by emission from unresolved radio galaxies. We have argued in this paper that radio galaxies -- active galaxies with mis-aligned jets -- are likely to also be the primary source of the astrophysical neutrinos observed by IceCube.  We have presented a simple physical model in which cosmic rays are confined by magnetic fields within radio galaxies for timescales of $t_{\rm esc} \sim 2.5 \, {\rm Myr} \times ({\rm PeV}/E_p)^{1/3}$, during which they scatter with gas to generate the observed diffuse fluxes of gamma rays and neutrinos. For cosmic rays accelerated by AGN with a spectral index of $\Gamma_p \simeq 1.8$, we can simultaneously accommodate the characteristics of Fermi and IceCube's observations, while also providing an attractive class of sources for the ultra-high energy cosmic rays. 

Smoking gun signals that would confirm the class of scenarios discussed here include the detection of neutrinos or ultra-high energy cosmic rays from indivdual, likely nearby, radio galaxies. From this perspective, the radio galaxy Centaurus A (Cen A) is particularly interesting.  At a distance of 3.8 Mpc, Cen A is the nearest radio galaxy, as well as the brightest at GeV energies (see Table 1 of Ref.~\cite{Hooper:2016gjy}). Furthermore, in 2010, the Auger Collaboration reported a modest excess of events above 55 EeV from directions within $\sim$$20^{\circ}$ of Cen A~\cite{Abreu:2010ab,Liu:2012sq}. Given the estimated 4\% chance probability of such an excess appearing randomly, however, further data will be required to confirm the authenticity of this signal.

Cen A is also a promising source for detection with future high-energy neutrino telescopes~\cite{Ahlers:2014ioa}. The gamma-ray flux from Cen A represents approximately 0.1\% of the total IGRB, and if we assume that the neutrino flux from Cen A is also equal to 0.1\% of IceCube's total astrophysical flux, we estimate that this source should generate a neutrino flux of $\sim 1.3 \times 10^{-9}$ GeV cm$^{-2}$ s$^{-1}$, which is a factor of $\sim$$20$ below the upper limits currently placed by the IceCube Collaboration~\cite{Aartsen:2013uuv}. Radio galaxies located in the northern hemisphere (including Cen B, NGC 6251, NGC 2484, 3C 264 and M 87) also represent promising targets for future point source searches with IceCube.








\bigskip
\bigskip

\textbf{Acknowledgments.} We would like to thank Tim Linden, Alejandro Lopez, Markus Ahlers, and Francis Halzen for helpful discussions. DH is supported by the US Department of Energy under contract DE-FG02-13ER41958. Fermilab is operated by Fermi Research Alliance, LLC, under Contract No. DE-AC02-07CH11359 with the US Department of Energy.

\bibliography{radio.bib}
\bibliographystyle{JHEP}

\end{document}